# Haunted House: A text-based game for comparing the flexibility of mental models in humans and LLMs

Brett Puppart, Paul-Henry Paltmann, Jaan Aru

Institute of Computer Science, University of Tartu

brett.puppart@gmail.com

**Abstract**

This study introduces "Haunted House[1]," a novel text-based game designed to compare the performance of humans and large language models (LLMs) in model-based reasoning. Players must escape from a house containing nine rooms in a 3x3 grid layout while avoiding the ghost. They are guided by verbal clues that they get each time they move. In Study 1, the results from 98 human participants revealed a success rate of 31.6%, significantly outperforming seven state-of-the-art LLMs tested. Out of 140 attempts across seven LLMs, only one attempt resulted in a pass by Claude 3 Opus. Preliminary results suggested that GPT o3-mini-high performance might be higher, but not at the human level. Further analysis of 29 human participants' moves in Study 2 indicated that LLMs frequently struggled with random and illogical moves, while humans exhibited such errors less frequently. Our findings suggest that current LLMs encounter difficulties in tasks that demand active model-based reasoning, offering inspiration for future benchmarks.

---

[1] The game can be played on the following link: https://paul-henryp.github.io/hauntedhouse/

# Introduction

The advent of transformer-based large language models (LLMs) has reignited the philosophical debate of human significance – a question that has persisted for millennia. Aristotle thought the function of humans was to live according to the rational principle, which was something that distinguished us from other animals (Aristotle, 2014). Back then, this might have seemed like a reasonable conclusion, as humans use complex language and abstract thinking to a degree that other animals simply do not. However, recent advancements in artificial intelligence (AI) are shining light on the possibility that in the future we might be living in a world in which our creation is more intelligent than us – or perhaps that this world is already here.

In many benchmarks comparing humans and AI, LLMs have shown a trend of rapid increase in performance. In SimpleBench, which measures common sense reasoning and social intelligence, GPT-4o scored only 17.8% and o1-preview 41.7% (Philip & Hemang, 2024). Recently, ARC-AGI – a long-standing benchmark for human-level intelligence – was beaten by OpenAI's GPT o3 (Chollet, 2024). Furthermore, scaled-up LLMs have been getting closer to the performance of domain experts. In the Frontier Math benchmark, in which solving a single problem takes hours of work from expert mathematicians (Glazer et al., 2024), previous LLMs had not exceeded an accuracy of 2%, whereas GPT o3 reached an accuracy of 25.2% (OpenAI, 2024). In Humanity's Last Exam, which consists of 3000 expert-level questions in domains such as mathematics, humanities, and natural sciences, GPT 4-o scored 3.3%, and GPT o1 scored 9.1% (Phan et al., 2025). This rapid improvement in performance raises the question of how intelligent these systems are in comparison to humans.

Empirical evidence suggests that LLM capabilities might not generalize in a way that human intelligence does. It has been shown that LLM performance is significantly worse with problems that sufficiently deviate from its training data (Wu et al., 2024). In one recent study, it was shown that o1-preview performance in a math dataset decreased by around 30% when the value of the variables was changed (Gulati et al., 2024). Similarly, Ullman's (2023) study showed that trivial alterations in Theory of Mind tasks significantly reduced LLM scores. This suggests that LLMs might not truly understand the underlying algorithm. Because if an agent understands the algorithm x to solve y, then small variations in y should not affect its performance.

To highlight LLMs lacking general intelligence, researchers have aimed to create tasks that are simple for humans but difficult for LLMs. It has been claimed that we will know that AGI is here when the exercise of creating such tasks becomes impossible (Chollet, 2024). Such benchmarks have measured pattern recognition (Chollet, 2019), common-sense reasoning (Philip & Hemang, 2024), and understanding of context (Levesque et al., 2012). However, to our knowledge, the ability to construct mental models has gotten little attention in previous tasks. We aim to fill this gap in the existing literature by designing a model-based reasoning task that is simple for humans but difficult for LLMs.

To reason about the world, humans construct mental models by compressing vast amounts of data into abstract internal representations (Johnson-Laird, 1983). These models allow us to predict the world and select the behavioral output that will most probably achieve our desired state (Jones et al., 2011). For example, if you are lost in a forest and come across a river, you might build two mental models: wandering in a random direction or following the river. Wandering aimlessly could leave you disoriented and further from help, whereas following the river increases your chance of finding a town and a way out. It has been argued that without building rich internal representations of the world, an AI system with human-level intelligence is impossible (Marcus, 2020).

There is a debate about whether such abstract representations could have emerged from next-token prediction in transformer-based LLMs (Mitchell, 2023). Confirming evidence comes from a study by Li et al. (2024) in which a transformer was trained to predict the legal moves of the game Othello, using only the sequences of game moves. The model developed an internal representation of the game state, which could be decoded from its activations (Li et al., 2024). Furthermore, it has been shown that color term embeddings in LLMs are relationally similar to perceptual color spaces (Abdou et al., 2021). This suggests that LLMs can capture certain aspects of human perceptual structure and the world through text-based training alone. However, it has been argued that such results might not generalize to the real world, which is a much more complex environment (Marcus, 2020; Mitchell, 2023).

Our mental models have to be flexible to dynamically adapt to changes in the environment (Jones et al., 2011). If there is a mismatch between new data and our existing mental model, we either fine-tune the parameters of the existing model or build an entirely new model – a process known as model updating (Filipowicz et al., 2016). Coming back to the example of being lost in the forest. You initially followed the river, but at some point, you get new data, as you hear people talking in some direction. You take this information into consideration, and you update your mental model, reaching the behavioral output: follow the sounds of the people to find your way out of the forest. In contrast, in most benchmarks, the task requires solving problems in which the variables remain constant.

In most benchmarks, the unwritten rule is that your current knowledge and the problem taken together either have sufficient information to solve it or not. In contrast, some problems in our everyday life require action in ambiguity to acquire sufficient information for solving the problem. You had to follow the river to hear the sounds of people, as your initial state did not include information about people's voices. There is some evidence that LLMs struggle with tasks that require active engagement with the problem to acquire new information. In Johri et al. (2025) study, LLMs had to diagnose diseases based on the given information. The results showed that LLM performance was worse when they had to collect information via active dialogue with another language model, compared to diagnosing based on a patient vignette. Furthermore, LLMs performed significantly worse when they were not given multiple-choice questions but had to conduct open-ended diagnoses (Johri et al., 2025).

In this article, we introduce a novel task called "Haunted House", designed to compare the flexibility of mental models in humans and LLMs. The task is in the format of a text-based game, in which the player must escape a haunted house while avoiding a ghost. The house contains nine rooms in a 3x3 grid layout. The players are required to build a mental model of the game environment one move at a time. They start with insufficient information for solving the problem, hence they must actively engage with the environment to collect new information. The variables in the house (e.g., the location of the ghost and the player) are changing dynamically, which requires the players to flexibly update their existing mental models as new information emerges. This task was designed with the explicit intent of being simple for humans but challenging for state-of-the-art LLMs. To test our task, we conducted two studies. In the first study, we collected data from 98 human participants to establish human performance. In the second study, we analyzed data from 29 additional human participants, focusing on their moves and errors to further understand human reasoning and compare their performance to that of LLMs. Lastly, we measured the effect of modified instructions on LLM performance.

**The task**

To assess model-based reasoning in LLMs and human participants, we designed a novel text-based game. In the game, players navigate a house with nine rooms in a 3x3 grid layout (rooms are labeled by columns A-C and rows 1-3, though players are not aware of this coordinate system). Players can move in the house by choosing their desired direction (left, right, up, down) (Figure 1). A legal move occurs when a room exists in the chosen direction. In other cases the move is considered illegal (e.g., trying to move right from the top-right corner). We set a maximum limit of 20 moves, counting both legal and illegal moves, to avoid LLMs getting stuck in endless loops of repetitive action. Before starting the game, players are required to read the instructions (Appendix A).

Players start in C1 (with the exit door), without the knowledge of their starting location. They must find the key hidden in one of the rooms, and return to the exit door to escape the house. They are guided by verbal clues they get as feedback each time they move (e.g., "The ghost is nearby," "The key is nearby," and "You cannot move there"). The full list of verbal clues and the conditions under which they are provided to the player are included in Appendix B. After finding the key and returning to the exit door, a clue informs the players that the exit door has relocated to the room of maximum distance from their current room (A3). After this, the ghost starts moving as the player reaches specific rooms in the house. The ghost movement algorithms are described below.

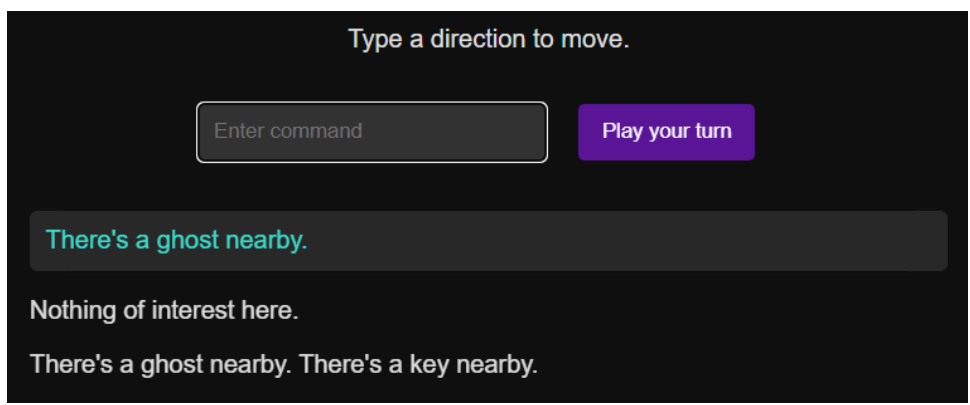

**Figure 1.** *The game as seen by the human participants.*

To complete the task, players need to reach the five following sub-objectives (Figure 2):

> 1. **Find the key in A1**. The players start in C1 with the exit door. Their first move is effectively random, either to B1 or C2. To solve this sub-objective correctly, the player has to visit both B1 and C2. In B1 the player is told, "The ghost is nearby. The key is nearby.". In C2 the clue "The ghost is nearby" is presented. If a player visits both of these rooms, then locating the key in A1 is a logical deduction because A1 is not adjacent to both C2 and B1. However, if the player's first move is to B1, they have to infer that moving back to C1 is a better choice than a 50% chance of finding the key and an equal chance of dying.

2. **Move back to C1**. After finding the key in A1, the player is presented with the clue, "You found the key! You will no longer be warned that the ghost is nearby." To solve this sub-objective correctly, they simply have to retrace their steps back to C1.

3. **Move to A2**. Reaching C1, the player is presented with the clue, "The layout of the house has changed. The door has moved to the location that is the maximum distance from your current room.". The player has to understand their location in the house and infer that the door has moved to A3. With the next move from C1 to either B1 or C2, the player is told that the ghost has moved one room down. To solve this sub-objective correctly, the player has to keep track of the ghost moving down and avoid going to B3.

4. **Avoid A3**. In A2, the player is told that the ghost has moved one room to the left. To solve this sub-objective correctly, the player has to keep track of the ghost moving to the left and blocking their way to the exit door. Thus, the player has to move back to either B2 or A2, where they are told that the ghost has moved two rooms to the right, freeing the exit door at A3.

5. **Move to A3**. Once the player reaches A3, the game is successfully passed.

**Figure 2.** *Visualization of one possible correct solution to the task.* The green circles and the numbers correspond to the player's location at a given point in time. The red circles correspond to the location of the ghost at a given point in time (e.g., the red circle in B3 says

"8-9" which means that when the player was in B1 (8) and A1 (9), the ghost was in B3; once the player moved to A2 (10), the ghost moved to A3 (10)).

# Study 1

The aim of Study 1 was to measure the average scores of humans and LLMs.

**Method**

*Participants and LLMs*. The invitations to participate in the study were shared on social media. There was no age limit for participating. There were in total 98 participants, of whom 65 were female (66%) and 34 male (34%). The average age was 24.2, ranging from 17 to 49. Six people completed the study in English and 92 in Estonian. As for LLMs used in this study, we picked the current state-of-the-art models to which we had access. The list includes GPT4-o, GPT 4-o mini, GPT o1-preview, Claude 3 Opus, Mistral Large 2411, Grok Beta, and Google Gemini 1.5 Pro. In a later study we also tested DeepSeek R1 and GPT o3-mini-high. We used the default sampling techniques, including top-p, top-k, and temperature, for all LLMs.

*Procedure*. Humans participated in the study on the LimeSurvey web platform. Because most participants were expected to be Estonians, we included the option of playing the game in Estonian as well as in English. Each human participant played the game only once. The game was programmed to fully automate data collection for human participants. Before beginning the task, they were required to read the game instructions (Appendix A). To collect data from LLMs, we engaged in an active dialogue with the models. We initiated each session with the prompt, "You are the player, solve the task…" followed by full instructions of the game. The LLMs generated their moves as text-based output. After each of their moves, we entered a prompt containing the corresponding verbal clues as feedback (Appendix B). Each model was tried 20 times in total.

**Results**

31 out of 98 human participants solved the task correctly, resulting in a 31.6% pass rate (Figure 3). Across seven LLMs, six did not complete the task in 20 attempts. Only Claude 3 Opus completed the task once in 140 total attempts by LLMs. As can be seen in Appendix C, the human pass rate was higher than LLMs for all levels of education.

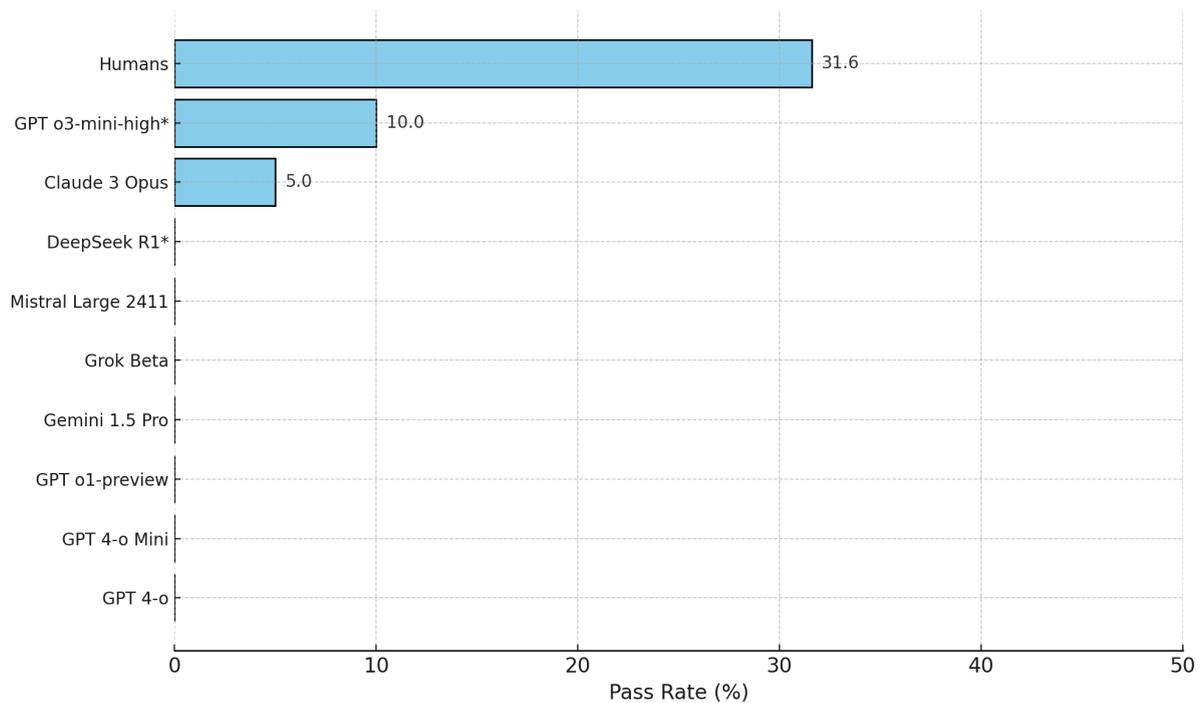

**Figure 3.** *Human and LLM performance in the Haunted House task.* Humans (n = 98). Each LLM had 20 attempts in total, except for models marked with * which had 10 attempts. The models marked with * were not part of Study 1 and were tried in a later study with modified instructions.

**Discussion**

The aim of Study 1 was to measure the average scores of humans and compare them to LLMs. The results established the average human pass rate at 31.6%, which far outperformed all LLMs. The performance of LLMs was very poor, as only one out of 140 attempts resulted in a pass by Claude 3 Opus. This suggested that LLMs struggle significantly more than humans in our task.

Our task was designed to be simple for the average human. However, the results suggested that humans also struggled in our task, as only about a third of the participants could solve the task correctly. As we did not collect data on the movement of human participants in Study 1, analyzing the causes of human failure was impossible. Additionally, we believed that analyzing and comparing LLM and human movement could provide further evidence that LLMs struggled with model building and updating. Because of this, we conducted an additional study in which we collected behavioral data on the movement of humans and compared them to LLMs.

# Study 2

The aim of study 2 was to analyze and compare human and LLM moves.

**Method**

*Participants and LLMs*. The invitations to participate in the study were shared in the forum of a university course on natural and artificial intelligence at the University of Tartu. There was no age limit for participating. There were in total 29 participants, of whom 10 were female (34%) and 19 were male (66%). The average age was 30.2, ranging from 11 to 47. Out of 29 participants, eight chose to play the game in English and 21 in Estonian. For LLMs, we used the same data we had collected in Study 1.

*Procedure*. Study 2 followed the same procedure as Study 1, using the same game and instructions. The only added step was recording the moves made by human participants. We aimed to compare human performance to LLMs in two metrics. Firstly, the amount of sub-objectives completed by both groups. The completion of sub-objectives provided valuable information about which parts of the game proved most difficult for humans and LLMs. Secondly, the amount of measurable errors made by both groups. Based on the moves alone, the following three types of errors could be identified:

1. **Failure to locate oneself**. Players committed this error when they made illegal moves (for which they got feedback: "You cannot move there") despite having enough information to know their exact location. For instance, if a player starts in the top-right corner (C1) and their first two moves are "up" and "right," they should recognize they are in the top-right corner of the house. If they subsequently make an illegal move, it implies they did not draw the correct inference about their location.

2. **Random guessing**. Before finding the key, three types of movement patterns indicated random guessing. The first one could be considered somewhat strategic. This occurs when a player starts in C1, moves "left" to B1, and then proceeds to A1 or B2. Although this might not seem like a mistake at first, it involves a random guess with a 50% chance of losing the game, making it an error in reasoning. Rather than relying on chance, a player could have returned to the safe position at C1. The second type involved illogical high-risk moves. This appeared when a player moved from C1 down to C2 and then chose B2 or C3 without having sufficient evidence of the location of the ghost. This choice did not improve their chances of finding the key but carried a 50% risk of immediate loss, making it a high-risk move with no compensating advantage. The third type was when a player was unaware of the fact that they had sufficient information to infer the location of the key and the ghost. This happened if they visited both B1 and C2 and still ended up walking to B2, where the ghost was located.

3. **Not tracking the ghost's movement**. This was recorded if, after finding the key and returning to C1, the player walked into the room with the ghost. It indicated that the player either did not deduce the ghost's location before finding the key or failed to update it after the ghost moved.

**Results**

Out of 29 human participants, nine (31%) finished the game without dying and in less than 20 moves. As can be seen in Table 1, 20 humans found the key in A1. Most LLMs failed before finding the key, with four finding the key at least once in 20 attempts. Based on the completion of sub-objectives, the worst-performing LLMs were GPT 4-o, GPT 4-o Mini and Gemini 1.5 Pro, as none of these models found the key in A1 in 20 attempts. The best-performing LLM in finding the key was o1-preview with a pass rate of 20%.

| Participant | Find the key | Move back to C1 | Move to A2 | Avoid A3 |
|---|---|---|---|---|
| Humans | 20 (69%) | 16 (55%) | 10 (34%) | 9 (31%) |
| GPT 4-o | 0 | 0 | 0 | 0 |
| GPT 4-o Mini | 0 | 0 | 0 | 0 |
| GPT o1-preview | 4 (20%) | 2 (10%) | 0 | 0 |
| Claude 3 Opus | 3 (15%) | 1 (5%) | 1 (5%) | 1 (5%) |
| Gemini 1.5 Pro | 0 | 0 | 0 | 0 |
| Grok Beta | 1 (5%) | 0 | 0 | 0 |
| Mistral Large 2411 | 1 (5%) | 0 | 0 | 0 |

**Table 1.** *Completion of sub-objectives for humans and LLMs.* Humans n = 29; LLM n = 20.

The most frequent error made by humans was random guessing, which was made by 18 participants out of 29 (62%). Nine of these random guesses were made on B1, where the players got the verbal clue "The ghost is nearby. The key is nearby.". Five made an illogical high-risk move from C2 to either B2 or C3. Four participants failed to infer the location of the ghost based on the verbal clues provided in C2 and B1. Tied for the second most frequent error was failure to locate oneself. Seven humans (24%) in total made moves that were evidence that they failed to infer their location and the boundaries of the house even after encountering sufficient information. Tied for the second most frequent error was the failure to update the location of the ghost. After getting back to C1, seven humans (24%) failed the game because they walked straight into the room that the ghost had just moved to.

All LLMs except Claude 3 Opus walked directly to B2 in at least 75% of the attempts. For analyzing errors, we picked Claude 3 Opus and GPT o1-preview because these were the most successful models in finding the key. GPT o1-preview made a random guess in a total of 17 attempts (85%). Out of these attempts, 11 were made from C2 and six from B1. In three attempts (15%), the model made an error in locating themselves in the house. Claude 3 Opus made a random guess in a total of 14 attempts (70%). Out of these attempts, eight were made from B1 and six from C2. In three attempts (15%), the model failed to locate themselves in the house.

Among the attempts, in which the player successfully found the key, GPT o1-preview made an error with locating itself in 75% and failed to track the ghost in 100% of the cases. Claude 3 Opus mislocated itself in 100% and failed to track the ghost in 75% of the cases. Humans mislocated themselves in 30% and failed to track the ghost in 55% of the cases.

Among the nine human participants who completed the entire task, seven did so without committing any measurable errors. One participant erred by randomly moving from B1 to A1 without first visiting C2, while another made a high-risk move from C2 to C3 that carried a 50% chance of failure. Once provided with sufficient information, none of these nine participants showed any measurable difficulty in orienting themselves within the house. In contrast, as shown in Appendix D, the sole LLM-based attempt that successfully completed the task was not free from such mistakes.

**Discussion**

The aim of Study 2 was to analyze and compare the moves made by humans and LLMs. We discuss the implications of these results in the "General Discussion" part of the article. The human pass rate in Study 2 was similar to the pass rate in Study 1, increasing the reliability of the original results.

Some humans struggled with understanding the instructions, as is evidenced by the feedback we got: "I think there is a bug. I found the key and went back to the start but the game still goes on;" "I couldn't see where to start and any grid;" "The instructions were a bit confusing, too large;" "I didn't notice that the exit door was in my starting room".

Two participants left us feedback, which hinted at a problem with the instructions. Namely, that it wasn't specified whether the ghost could move in the house on its own. This could have reduced the validity of the results. Therefore, we aimed to measure the effect of this potential confounding variable.

## Investigating the effect of modified instructions

**Method**

To investigate whether confounding variables affected the validity of our results, we conducted additional trials with modified instructions to address specific criticisms about our methodology. For these trials, we selected the two of the best-performing LLMs from our initial studies – Claude 3 Opus and GPT o1-preview. Additionally, we included DeepSeek R1, an open-source LLM with high performance in many benchmarks (DeepSeek-AI et al., 2025), and GPT o3-mini-high, a newly released model by OpenAI (OpenAI, 2025). We tested the models under two sets of modified instructions. For each of the two criticisms, we conducted ten additional trials with Claude 3 Opus and GPT o1-preview. Given time constraints, DeepSeek R1 and GPT o3-mini-high were only tested with the "Ghost" instructions as they more closely resembled the ones used in our original studies.

One concern from our original task was that it was not explicitly stated whether the ghost could move in the house. This ambiguity may have led to worse performance from LLMs, as the models could have mistakenly inferred that, after making their first move and receiving the verbal clue "The ghost is nearby," the ghost could have moved to their starting room. Thus, they might have considered their second move as a random guess with an equal chance of failure in each direction. To control whether this affected LLM performance, we added the information to the instructions that the ghost remains stationary unless stated otherwise. We refer to these instructions with the name "Ghost" (Appendix A).

Another concern was that perhaps humans were simply more acquainted with the choice of movement (e.g., left, right), and some other method would have generated better performance from LLMs. To control this, we kept the previous addition about the ghost's movement in the instructions. We further added a description of the coordinate system and explicitly stated that they start in room C1. To move in the house, LLMs had to type the coordinates of the rooms. This modification allowed for further analysis of whether LLM poor performance was caused by their inability to assign proper coordinates to rooms or their more general inability to map the house and its contents. We refer to these instructions with the name "Coordinates" (Appendix A).

**Results**

The "ghost" instructions did not increase the performance of GPT o1-preview and Claude 3 Opus (Figure 5). However, the newer models DeepSeek R1 and GPT o3-mini-high showed promising preliminary results, as both of these models found the key in at least 10% of the attempts. GPT o3-mini-high passed the entire task in one attempt with no measurable errors, suggesting that it might be the state-of-the-art model in the Haunted House task.

The "Coordinates" instructions improved LLM performance in finding the key but not in passing the game. Both Claude 3 Opus and GPT o1-preview found the key in at least half of the attempts. However, out of 20 total attempts, only one by o1-preview resulted in passing the entire game. Closer inspection revealed that this passing attempt was not free of errors, as the model used random guessing to find the key.

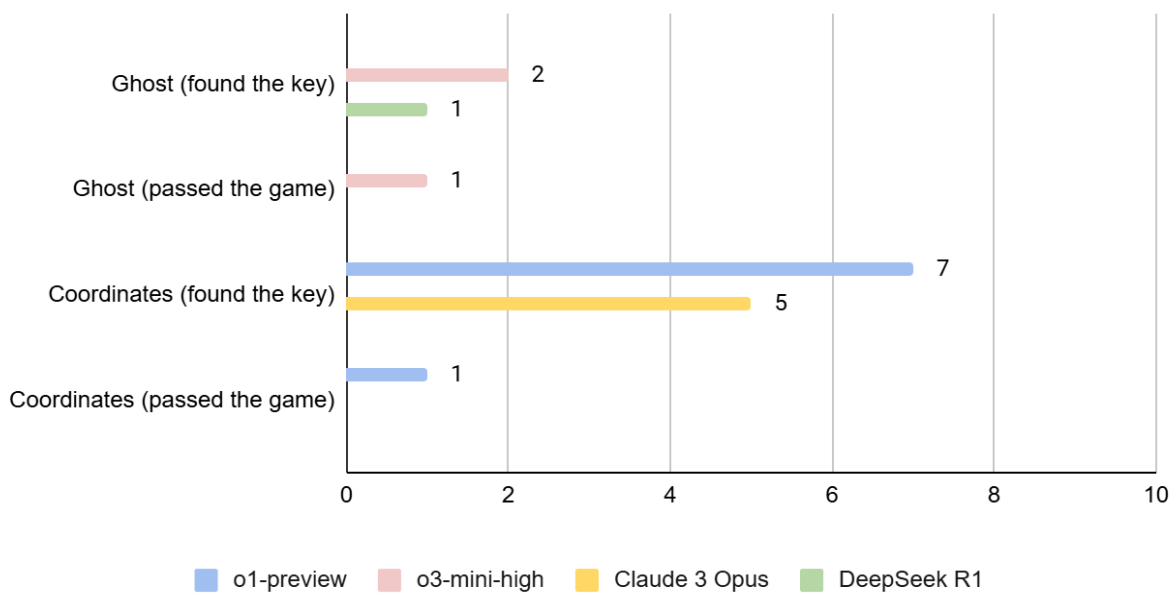

**Figure 5.** *LLM performance with modified instructions.* "Ghost" stands for the added instruction that the ghost stays stationary unless stated otherwise. "Coordinates" stands for the added instructions about moving via coordinates instead of directions. In total, there were 10 attempts for each LLM. The performance of DeepSeek R1 and GPT o3-mini-high was measured only under the "Ghost" instructions.

**Discussion**

The modified instructions did not increase LLMs performance in passing the game. However, analyzing the reasoning process of DeepSeek R1 and GPT o3-mini-high provided valuable insight into their struggles with the task.

While these models were great at hypothesis generation, they encountered substantial difficulties in selecting the correct behavioral output. At times, they considered hypotheses that humans are very unlikely to consider. For example, DeepSeek R1 entertained the possibility that the house was toroidal, and o3-mini-high considered that perhaps there was a "clue hierarchy" in which the clue that was presented first hinted at the first row and the clue presented secondly hinted at the second row. In most attempts, after making their first legal move, these models considered the correct move of returning to the starting room to avoid the ghost. However, this move was frequently abandoned for seemingly irrational reasons. For example, DeepSeek R1 made its first legal move to B1, where it got the clue "The ghost is nearby. The key is nearby.". It correctly identified C1 as a safe move but went on to reason, "But since the clue says both are nearby, maybe moving in one direction leads to the key, another to the ghost. So perhaps I have a 50% chance if there are two possible directions. Wait, but there are three directions: left, right, down. But right leads back to starting room, which is safe but no key. So maybe left and down are the options". Based on this line of reasoning, the model went on to move to B2 and encountered the ghost.

Additionally, both models used techniques to map the house but frequently became disoriented and lost. They attempted to assign coordinates to visited rooms (e.g., starting room = 1/3 etc.), yet often reassigned coordinates randomly or had general difficulty with understanding the spatial dimensions of the house. For example, after making its first moves to "up" and "right", o3-mini-high inferred that it was in the bottom left corner of the house. Its first legal move was to the right, after which it reasoned that the starting room was to the left. Here is an excerpt of the o3-mini-high reasoning process: "Moving right to the starting room seems safest without jeopardizing the key. Choosing between left and down is tricky, given the risk of encountering the ghost".

# General Discussion

The objective of this study was to develop a model-based reasoning task that would be simple for humans and difficult for state-of-the-art LLMs. Humans far outperformed all LLMs, signaling their superiority in our task that requires mental model updating and active engagement. On average, 31.6% of humans completed the task, in contrast to the best-performing LLM, Claude 3 Opus, which achieved a pass rate of 5%. From a total of 140 attempts across seven LLMs, only one attempt resulted in a pass. Preliminary results from GPT o3-mini-high and DeepSeek R1 showed promising results. GPT o3-mini-high scores suggest that it reached state-of-the-art performance in our task, solving the entire task in 10% of the attempts – which is still not on the human level.

Despite being designed as a simple task for humans, the results indicated that the task was not as simple for the average human as we initially predicted. In many existing benchmarks with a similar goal as our task, the human average is usually a lot higher. In ARC-AGI, the human benchmark is 85% (Chollet, 2024), and in SimpleBench it is 83.7% (Philip & Hemang, 2024). It is possible that the amount of instructions and information provided by verbal clues was too much to follow for humans, whose working memory capacity is limited (Miller, 1956; Wilhelm et al., 2013). The feedback we got from human participants in Study 2 highlights that at least some of them found the instructions too long or complex to follow. Future model-based reasoning tasks should aim to decrease the amount of instructions to decrease the burden on human working memory.

Another explanation for low human performance might be linked to motivation. There is empirical evidence that motivational factors affect human performance in reasoning (Dawson et al., 2002). Humans have a tendency to avoid analytical thinking to save cognitive resources (Kahneman, 2011). Due to this, the average scores of humans in cognitive tasks might be skewed by cognitive tendencies and not measure cognitive capability *per se*. Some humans in our task might have lacked the motivation to build a mental model. Therefore, it is possible that human scores in our task would have been higher in a controlled environment in which the problem solver was rewarded for a correct solution. Another possibility would have been to force participants to confirm their understanding of the task by answering a few questions about the details before they could move on to playing the game.

In contrast, the poor performance of LLMs is best explained by their inability to construct working mental models of the game environment. The first supportive evidence for this claim is that even in the attempts in which LLMs completed sub-objectives of the game, they made very simple errors. For example, o1-preview found the key in 20% of the attempts. After finding the key, its next sub-objective was to simply retrace its steps back to the starting room, which o1-preview could only do in half of these attempts. Furthermore, Claude 3 Opus and o1-preview found the key in a total of seven attempts, out of which six (86%) contained illegal moves that were evidence of their failure to locate themselves after encountering sufficient information to do so. In contrast, only 30% of the humans made such positioning errors in the attempts in which they found the key.

The second supportive evidence is that the explanations of LLMs did not correspond to the variables and the requirements of the task. There were cases where the model was just one step away from the key but chose a nonsensical move, e.g., GPT 4-o Mini stating, "Let's try down then, as it seems like our only other choice," when moving left would have reached the key. Claude 3 Opus made a similar error and provided a nonsensical explanation: "If the room to the left contained the key, the clue in the starting room likely would have said "the key is nearby", but it didn't.". Even GPT o3-mini-high and DeepSeek R1 explanations revealed many errors. Although these models tried building mental models by assigning coordinates to visited rooms (e.g., 1/1 and 2/2), the coordinates were sometimes seemingly randomly reassigned, which caused the models to err. There were many instances of the models getting disoriented in the house, even after the very first move. Reasoning models proved to be good at mental model generation, mostly considering the correct solutions in their reasoning process but ultimately disregarding them for illogical reasons. Furthermore, the performance of LLMs revealed many instances of hallucination, which is in accordance with previous work highlighting LLM hallucinations (Borji, 2023). For example, Claude 3 Opus hallucinated an "unknown room" and tried to move there (Appendix D). Gemini 1.5 Pro thought it was between walls, which was impossible in our task.

The third supportive evidence is that solving a simplified version of the task did not increase LLM performance. Initially, the task was designed to be more difficult for LLMs by withholding information about their starting location from the player. To assess whether this contributed to their poor performance, we conducted additional trials in which the player moved by writing the coordinates of the rooms (e.g., A1, B2, C3), and their starting location was given to them in the instructions. Despite this modification, LLM performance did not improve, with only one attempt out of 20 successfully finding the key. The continued failure under these conditions suggests that LLMs struggled not only with mapping coordinates to rooms but also with integrating and updating key spatial details – such as the positions of the key, the player, and the ghost – within the existing coordinate framework.

It has been shown that LLMs can build internal representations that they were not specifically trained for (Li et al., 2024) and that their embeddings can correspond to perceptual structures in the real world (Abdou et al., 2021). Our findings suggest that such capabilities might not generalize to problems that the LLM has not encountered before. Future research should continue designing tasks that are simple for humans but difficult for LLMs to highlight their limitations. Additionally, more robust and large-scale benchmarks are needed to systematically assess the extent of LLM difficulties with model-based reasoning.

**Acknowledgments**

We are grateful to Marharyta Dominich for her helpful comments. This work was supported by the Estonian Research Council grants PSG728 and Tem-TA 120 and the Estonian Centre of Excellence in Artificial Intelligence (EXAI), funded by the Estonian Ministry of Education and Research.

44

## Appendix A - Complete instructions of the game

The following instructions are the ones provided to the players in Study 1 and 2. The footnotes describe how the modified instructions "Ghost" and "Coordinates" differed from the original ones.

Welcome to the Haunted House game! 👻

In this text-based game, you are trapped in a house with an evil ghost. The exit door is locked. You have to find the key, unlock the door, and leave the house. You must do this in less than 20 moves. Please read the following instructions carefully!

Here's the information that you have:[2]

- The house is a 3x3 grid with 9 rooms in total.[3]
- You do not know which room you start in. [4]
- The locked exit door is in your starting room.
- You must avoid the ghost - if you enter the same room, the game ends.

Movement:

- You can move one room at a time.
- You cannot move diagonally.
- To move, simply type the direction you want to go (left, right, up, down). [5]

Clues:

- When you enter a new room, you will receive a verbal clue.
- If the clue says "The key is nearby", it means that your next move can lead to the room with the key.
- If the clue says "The ghost is nearby", it means that your next move can lead to the room with the ghost.
- There are many other verbal clues so pay close attention to them!

To start the game, type the first direction you want to move.[6]

---

[2] In "Ghost" and "Coordinates" there was an additional bullet point in this subsection: "The ghost remains stationary in the house unless stated otherwise".
[3] This was replaced in "Coordinates" with "The house is a 3x3 grid where the columns are labeled A, B, and C from left to right, and the rows are numbered 1, 2, and 3 from top to bottom. For example, the top-left room is A1 (column A, row 1), and the bottom-right room is C3 (column C, row 3)".
[4] This was replaced in "Coordinates" with "You start in C1".
[5] This was replaced in "Coordinates" with "To move, simply type the coordinates of the room you want to go (e.g. A2, C3, B1)".
[6] This was replaced in "Coordinates" with "To start the game, type the first room you want to move to".

**Appendix B - List of verbal clues**

C1 "There's nothing of interest here"

C2 "You cannot move there"

C3 "There's a ghost nearby"

C4 "There's a key nearby"

C5 "You found the key! You will no longer be warned that the ghost is nearby"

C6 "The layout of the house has changed. The door has moved to the location that is the maximum distance from your current room"

C7 "The ghost has moved one room down"

C8 "The ghost has moved one room left"

C9 "The ghost has moved two rooms right"

C10 "Congratulations - You have escaped the haunted house!"

C11 "Game over - You encountered the ghost!"

**Appendix C. Participants' levels of education in Study 1**

| Education | n | Pass rate (%) |
|---|---|---|
| Completed university degree | 18 | 16.7 |
| Currently enrolled in university | 43 | 34.9 |
| Completed high-school | 25 | 32 |
| Currently enrolled in high-school | 12 | 41.7 |

## Appendix D. Claude 3 Opus passing attempt

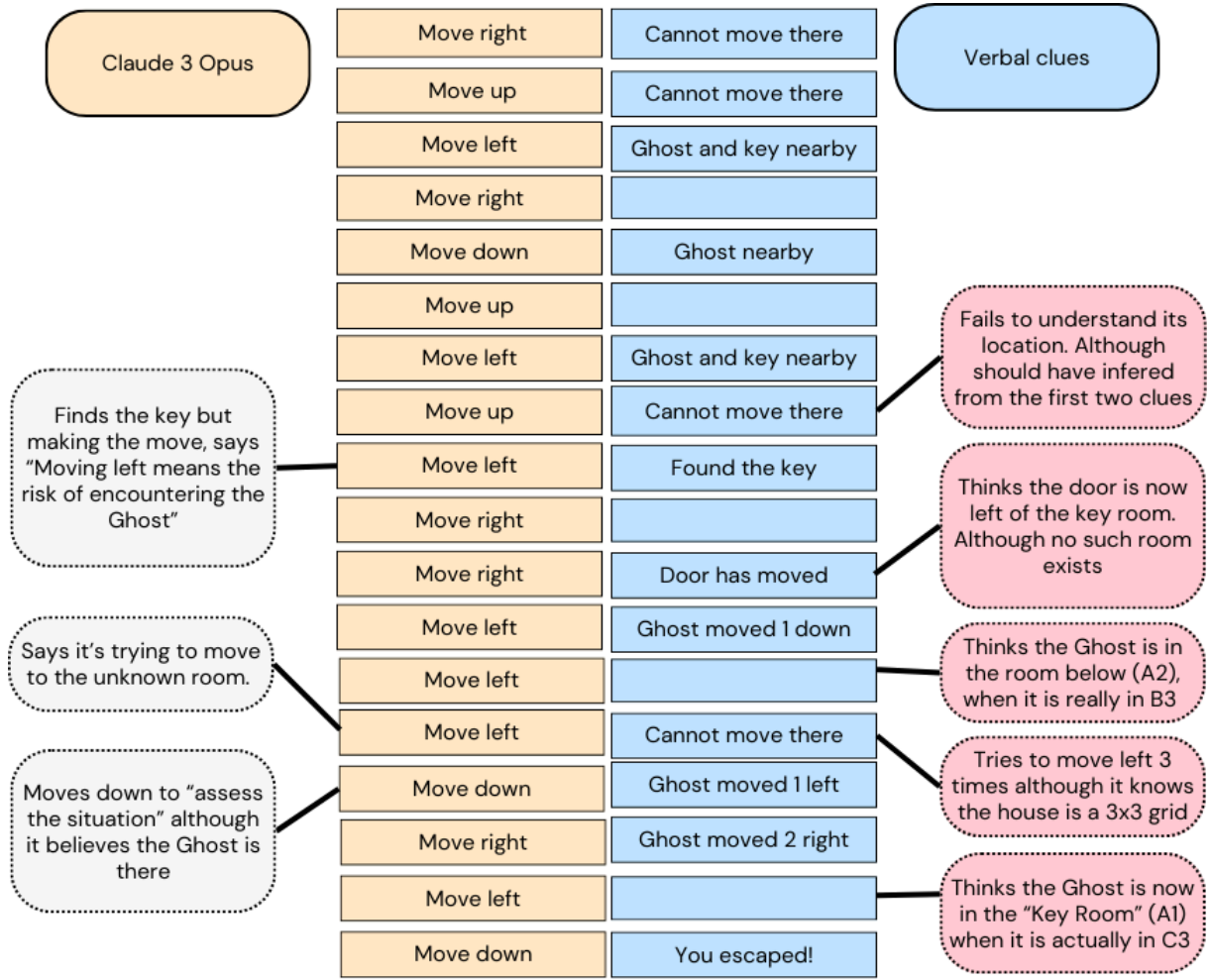

**Appendix D.** *Claude 3 Opus passing attempt.* The red rounded squares are mistakes made by the model. The grey rounded squares are additional information provided by the model.